\documentclass[runningheads,a4paper]{llncs}

\usepackage{amssymb}
\usepackage{graphicx}

\usepackage{url}
\newcommand{\keywords}[1]{\par\addvspace\baselineskip
\noindent\keywordname\enspace\ignorespaces#1}

\begin{document}

\mainmatter  

\title{Out-of-equilibrium dynamics in systems with long-range interactions: characterizing quasi-stationary states}
\titlerunning{Characterizing quasi-stationary states}

\author{Pierre de Buyl}
\authorrunning{P. de Buyl}

\institute{Center for Complex Systems and Nonlinear Phenomena\\
Universit\'{e} libre de Bruxelles - CP231\\
Av. F. Roosevelt, 50\\
B-1050 Brussels}

\maketitle

\begin{abstract}
  Systems with long-range interactions (LRI) display unusual thermodynamical and
  dynamical properties that stem from the non-additive character of the
  interaction potential. We focus in this work on the lack of relaxation to
  thermal equilibrium when a LRI system is started out-of-equilibrium. Several
  attempts have been made at predicting the so-called quasi-stationary state
  (QSS) reached by the dynamics and at characterizing the resulting transition
  between magnetized and non-magnetized states. We review in this work recent
  theories and interpretations about the QSS. Several theories exist but none of
  them has provided yet a full account of the dynamics found in numerical
  simulations.

  \keywords{Vlasov equation ; long-range interactions}
\end{abstract}

\section{Introduction}

Systems with long-range interactions (LRI) display unusual thermodynamical and
dynamical properties such as ensemble inequivalence, lack of relaxation to
equilibrium or broken ergodicity (see
Refs.~\cite{campa_et_al_phys_rep_2009,Bouchet:2010p3333} for a review of the
field). These properties stem from the non-additive character of the
interparticle interaction potential. Let us mention a few systems belonging to
the LRI class: gravitational systems, non-neutral plasmas, 2D fluid dynamics,
etc.

We focus in this work on the lack of relaxation to thermal equilibrium in LRI
systems when the system is initiated in an out-of-equilibrium state. This
phenomenon leaves the system in an intermediary stage of the dynamics that is
called a quasi-stationary state (QSS)
\cite{yamaguchi_et_al_physica_a_2004}. This state does not correspond to the
equilibrium predicted by statistical mechanics and its lifetime increases
algebraically with the number $N$ of interacting particles in the system.

The occurrence of QSS should be taken into account if one is interested in the
actual properties of a system. The time needed to reach thermal equilibrium may
prevent a proper observation of equilibrium properties in the available
experimental or simulational setting.

In this work, we review the generic steps of the out-of-equilibrium dynamics of
LRI systems and use the paradigmatic Hamiltonian Mean-Field (HMF) model and its
Vlasov formulation to illustrate those steps. Then, several theories attempting
to predict or describe the QSS are reviewed.



\section{The Hamiltonian Mean-Field model and the Vlasov equation}

Let us consider the Hamiltonian Mean-Field (HMF) model introduced by Antoni and
Ruffo~\cite{antoni_ruffo_1995}. This model aims at reproducing the collective
behavior of more complex models with ferromagnetic or gravitational
interactions, for instance.

The particles in the HMF model lie in a $1$-dimensional periodic space with
position $\theta \in [-\pi:\pi[$. The $N$-body Hamiltonian is
\begin{equation}
  \label{eq:HMF}
  \mathcal{H} = \sum_{i=1}^N \frac{p_i^2}{2} + \frac{1}{2N} \sum_{i,j=1}^N 
  \left( 1 -\cos (\theta_j-\theta_i)\right)
\end{equation}
where $\theta_i$ is the position (in $[-\pi:\pi[$) of particle $i$ and $p_i$ is
its momentum. $N$ is the total number of particles in the system.

One may also consider the continuum limit of the HMF model. This leads to the
Vlasov equation
\begin{equation}
  \label{vlasov}
  \frac{\partial f}{\partial t} + p \frac{\partial f}{\partial \theta} - 
  \frac{\partial V[f](\theta,t)}{\partial \theta}
  \frac{\partial f}{\partial p} = 0~,
\end{equation}
where
\begin{equation}
  \label{potential}
  V[f](\theta,t)=\int d \theta' dp' f(\theta',p',t)\left(1-\cos (\theta'-\theta)\right)~,
\end{equation}
is the interaction potential.

The mean field, or magnetization,
\begin{equation}
  {\mathbf m}=\frac{1}{N} \sum_{i=1}^N (\cos \theta_i, \sin \theta_i)=(m_x,m_y)~,
\end{equation}
is used to follow the dynamical evolution of the HMF model. In the continuum
limit.
\begin{equation}
  {\mathbf m}=\int d\theta~dp~ f(\theta,p) (\cos \theta, \sin \theta)=(m_x,m_y)~,
\end{equation}
The norm of $\bf m$ is denoted $m$.

Equilibrium statistical mechanics allows one to compute the value of $m$
for a given energy or temperature. The authors of Ref.~\cite{antoni_ruffo_1995}
observed a discrepancy in the caloric curve between the theory and their
simulations, close to the second order transition separating the magnetized
phase ($m>0$) from the homogeneous phase ($m=0$). Further
investigations revealed the origin of the discrepancy: the system of particles
had not reached thermodynamical equilibrium in the simulations.

The evolution of many systems with long-range interactions consists in the
following generic steps:
\begin{enumerate}
\item An initial condition that does not correspond to equilibrium.
\item Violent relaxation: the observables in the system undergo strong
  changes. The time scale of this step does not depend on the number of
  particles $N$.
\item Quasi-stationary state (QSS). This state may either be stationary or present
  oscillations. Its lifetime grows algebraically with $N$.
\item Equilibrium: the state that is predicted by equilibrium statistical
  mechanics.
\end{enumerate}
Simulations illustrating this dynamical evolution may be found in
Refs.~\cite{yamaguchi_et_al_physica_a_2004,yamaguchi_pre_2008}, for instance.
An important consequence of the occurrence of QSS is that in addition to the
equilibrium phase transitions that the system may experience, one has to
consider an ``out-of-equilibrium phase diagram'' that is based on the
magnetization found in the QSS. In the thermodynamic limit $N\to\infty$, this
``out-of-equilibrium phase diagram'' is the relevant one.

\section{Theories for the quasi-stationary states}

Several attempts have been made at predicting the quasi-stationary state reached
by the dynamics. We review several of those attempts, namely: Lynden-Bell's
theory that is based on an entropy maximization principle
\cite{lynden-bell_1967,antoniazzi_et_al_prl_2007}, the exact stationary regime
theory proposed by de Buyl, Mukamel and Ruffo
\cite{de_buyl_et_al_self-consistent_pre_2011} and a dynamical reduction proposed
by Levin~\cite{pakter_levin_prl_2011}.

None of the aforementioned theories is able to take into account the existence
of states whose observables are not constant in time, i.e. they predict a time
independent distribution $f(\theta,p)$. It is known that dynamical resonances
lead to oscillating regimes~\cite{antoni_ruffo_1995,antoniazzi_califano_prl} and
those cannot be predicted.

\subsection{The theory of Lynden-Bell}

In 1967, Lynden-Bell~\cite{lynden-bell_1967} devised a theory to compute the
relaxed state of gravitational systems obeying a Vlasov equation. His theory is
based on the maximization of an entropy functional that takes into account the
incompressible character of the distribution function in Vlasov dynamics. The
computation is based on a coarse graining of phase space but leads to a
continuous prediction for the distribution function.

Lynden-Bell's theory has been applied with success to the prediction of the
intensity of the Colson-Bonifacio model for the free-electron
laser~\cite{barre_et_al_pre_2004} and to provide an out-of-equilibrium phase
diagram for the HMF model~\cite{lynden-bell_1967,antoniazzi_et_al_prl_2007}.

\subsection{BGK like theory}

Based on the fact that a distribution function that only depends on the energy
is stationary in Vlasov dynamics, one may try to construct stationary
states. This approach is well know in plasma physics as Bernstein-Greene-Kruskal
modes~\cite{bgk_1957}. The authors of
Ref.~\cite{de_buyl_et_al_self-consistent_pre_2011} develop this idea while
proposing an approximate correspondence between the initial condition and the
state that is reached by the system.

The distribution $f(\theta,p)$ is expressed directly as a function of the energy
distribution function of the initial condition. For low values of the initial
magnetization, the theory fails to predict the final magnetization. Else, it
provides good results and predicts a second-order phase transition for $\langle
m \rangle$ and $\langle m_x \rangle$. This theory is based purely on dynamical
consideration and as such provides interesting complementary information with
respect to Lynden-Bell's theory.

\subsection{Core-halo and envelope}

The authors of Ref.~\cite{pakter_levin_prl_2011} propose an ansatz for the
distribution function $f$ that reproduces the core-halo structure found in the
phase space of the HMF model
\begin{equation}
  f_S(\theta,p) = \eta_0 \left[ \Theta(\epsilon_F - \epsilon) + \chi~\Theta(\epsilon_h - \epsilon) ~ \Theta(\epsilon - \epsilon_F) \right] ~,
\end{equation}
where $\epsilon(\theta,p) = p^2/2 + (1- M_S\cos\theta)$, $M_S$ is the value of
the magnetization and $\Theta$ is the Heaviside function.

This ansatz requires the determination of the energy levels ($\epsilon_F$ for
the core and $\epsilon_h$ for the halo) and of the magnetization $M_S$. Those
values are provided by a reduced dynamical equation. This theory is tested on
the transition between magnetized and homogeneous regimes in the HMF model and
predicts a first-order like transition for $\langle m \rangle$ and $\langle m_x
\rangle$. The order of the transition is confirmed by simulation data for
$\langle m_x \rangle$. Simulation data for $\langle m \rangle$ is not given
however.

\section{Discussion and conclusion}

Out of the existing theories aimed at predicting the quasi-stationary states
(QSS) that have been applied to the Hamiltonian Mean-Field model, none is able
to predict the regimes in which oscillations are found. As is pointed out in
Ref.~\cite{de_buyl_et_al_cejp_2012}, one may relate the time averages of the
squared norm of the magnetization to the one of the $x$ component of the
magnetization\footnote{Here, $m_y$ can be set equal to zero without loss of
  generality.} by the following relation:
\begin{equation}
\langle m^2 \rangle=\langle m_x^2 \rangle= \langle m_x \rangle^2 +\sigma_{m_x}^2~,
\end{equation}
where $\sigma_{m_x}$ is the time-wise standard deviation of $m_x$. The choice of
an observable thus impacts the results that is found in simulations for
non-steady QSS, explaining the different results between
Ref.~\cite{antoniazzi_et_al_prl_2007} and Ref.~\cite{pakter_levin_prl_2011}. As
soon as the QSS displays oscillations in the magnetization, $\sigma_{m_x}^2 > 0$
and $\langle m^2 \rangle \ne \langle m_x \rangle^2$. The phase diagram provided
by Lynden-Bell's theory \cite{antoniazzi_et_al_prl_2007} still represents the
most ensemble view of the QSS for the HMF model as well as an actual
interpretation in terms of phase transitions.

We have reviewed in this work recent advances in the understanding of the
out-of-equilibrium dynamics in systems with long-range interactions. Several
theories exist but none of them has provided yet a full account of the dynamics
found in numerical simulations. Progress in this direction has been made by the
construction of counter-rotation BGK clusters by
Yamaguchi~\cite{yamaguchi_pre_2011}. This construction is however not
predictive.

\subsubsection*{Acknowledgments.} The author would like to acknowledge
interesting discussions and collaborations with R. Bachelard, G. De Ninno,
D. Fanelli, P. Gaspard, D. Mukamel and S. Ruffo.

\end{document}